\documentclass{aastex63}

\received{2019 Aug 26}
\revised{2019 Oct 17}
\accepted{2019 Oct 30}
\submitjournal{Icarus}

\shorttitle{Dust Devil Radii and Heights}
\shortauthors{Jackson}

\usepackage{amsmath}
\usepackage{amssymb}

\begin{document}

\title{On the Relationship between Dust Devil Radii and Heights}

\correspondingauthor{Brian Jackson}
\email{bjackson@boisestate.edu}

\author[0000-0002-9495-9700]{Brian Jackson}
\affiliation{Boise State University, Dept.~of Physics\\
1910 University Drive, Boise ID 83725-1570}



\begin{abstract}
The influence of dust devils on the martian atmosphere depends on their capacity to loft dust, which depends on their wind profiles and footprint on the martian surface, i.e., on their radii, $R$. Previous work suggests the wind profile depends on a devil's thermodynamic efficiency, which scales with its height, $h$. However, the precise mechanisms that set a dust devil's radius have remained unclear. Combining previous work with simple assumptions about angular momentum conservation in dust devils predicts that $R \propto h^{1/2}$, and a model fit to observed radii and heights from a survey of martian dust devils using the Mars Express High Resolution Stereo Camera agrees reasonably well with this prediction. Other observational tests involving additional, statistically robust dust devil surveys and field measurements may further elucidate these relationships.
\end{abstract}


\section{Introduction} \label{sec:introduction}
The martian atmosphere is dusty -- analyzing spectra collected by Mars Global Surveyor (MGS) Thermal Emission Spectrometer (TES), \citet{2004Icar..167..148S} found globally averaged dust infrared optical depths $\tau$ often exceed 0.15, comparable to the daily smog layer in Los Angeles \citep{2007JGRD..11222S21R}, and large dust storms can drive $\tau$ to $\gg$ 1 \citep{2002Icar..157..259S}. The suspended aerosols absorb and scatter radiation, modifying the atmospheric heat budget. \citet{2002Icar..157..259S} estimated Mars' 2001 global dust storm drove atmospheric temperatures up by at least 40 K, and the perpetually suspended background haze provides warming of $\sim$ 10 K \citep{2004JGRE..10911006B}. Dust removal/deposition varies regionally \citep{2006JGRE..111.6008K}, and fluctuations in polar deposition could alter the cap albedo and sublimation \citep{1995JGR...100.5501H}. Thus, the dust cycle is intimately woven into the fabric of Mars' climate.

The martian dust cycle is driven, in part, by dust devils, convective vortices rendered visible by dust. At the core of a dust devil, surface heating results in positive temperature and negative pressure excursions, which fall off with radial distance. The buoyant air ascends to roughly the top of the planetary boundary layer \citep{2015Icar..260..246F}, where the dust may be carried away by regional winds. Meanwhile, near the surface, surrounding air is drawn in, conserving vorticity and giving a tangential wind field at a devil's eyewall. 

Although devils clearly contribute to the atmospheric dust budget on Mars, their exact contribution remains highly uncertain. Based on imagery collected by the Spirit rover on Mars, \citet{2006JGRE..11112S09G} estimated that devils contribute only a tenth as much atmospheric dust as regional dust storms. A survey involving space-based imagery estimated devils are an important but perhaps not dominant source of dust \citep{2006JGRE..11112002C}. And \citet{2016SSRv..203...89F} suggested dust devils may contribute as much as 75\% of the total dust flux to the martian atmosphere. 

Key to resolving this uncertainty is an accurate assessment of the martian dust devil population and its dust-lifting potential. In this vein, ground-based surveys using the meteorological instruments on-board landers provide a powerful tool. These surveys involve sifting pressure time-series for the short-lived, negative pressure excursions that arise when a convective vortex passes near the lander \citep[e.g.][]{Ellehoj_2010, 2018Icar..299..308O}. These surveys have several advantages -- pressure time-series are often collected throughout the martian day, allowing for more accurate occurrence rate estimates; and they probe the internal structures of dust devils, providing important tests for physical models \citep{2000JGR...105.1859R}. However, these surveys may suffer from complex bias and selection effects \citep{2018Icar..299..166J}. Also, since the required wind speed data are almost always lacking, it is impossible to directly estimate the devils' physical sizes, required to estimate the areas over which devils lift dust and therefore their dust-lifting. On the other hand, space-based imaging surveys allow assessment of dust devil sizes and dust-lifting \citep{2006JGRE..11112002C}, but image resolution usually limits detections to the largest and least common devils \citep{2009Icar..203..683L}. Moreover, the images alone reveal little to nothing regarding the devils' internal structure, pressure, temperature, and wind profiles.

To bridge this gap, I adapt previously developed thermodynamic models for dust devils, supplemented by simplified assumptions regarding their angular momenta, to derive scaling relations between dust devil radii, pressure profiles, wind speeds, and heights. The relations predict, for example, that the radius scales with the square root of a devil's height. They also predict how radius depends on environmental conditions such as wind shear and atmospheric scale height. To check this model, I compare the radius-height scaling to data from the imaging survey reported in \citet{2008Icar..197...39S} and find reasonable agreement. Finally, I discuss possibilities for future work.

\section{Model} \label{sec:model}
For the analysis here, I assume a dust devil consists of a small, steady-state convective plume with a radial pressure structure resembling a Lorentz profile and a velocity structure resembling a Rankine vortex \citep{2016SSRv..203..209K}. The eyewall of the dust devil occurs at the peak in the velocity profile at a well-defined distance $R$ from the convective center. Far from the dust devil center, the wind field carries angular momentum inward along horizontal flowlines. Turbulent drag along the surface dissipates some (but not all) of the mechanical energy, providing the frictional dissipation required to establish a steady-state \citep{1998JAtS...55.3244R}. Decades of field work corroborate this model in broad strokes \citep[e.g.][]{2016SSRv..203...39M}, but statistically robust and detailed in-situ measurements of active dust devil structures remain undone.

At the dust devil's eyewall, cyclostrophic balance applies, and the pressure gradient force balances the centrifugal force:
\begin{equation}
    \dfrac{1}{\rho}\left( \dfrac{dp}{dr} \right) = \dfrac{\upsilon^2}{R},
\end{equation}
where $\rho$ is the atmospheric density near the surface, $p$ the pressure, $r$ radial distance from the devil's center, and $\upsilon$ the tangential velocity. The pressure structure follows a Lorentz profile:
\begin{equation}
    p(r) = p_{\infty} - \dfrac{\Delta p}{1 + \left( r/R \right)^2},\label{eqn:pressure_profile}
\end{equation}
where $p_{\infty}$ is the ambient pressure, and $\Delta p$ is the depth of the pressure perturbation at the devil's center. Calculating the pressure gradient from this profile and equating it to the centrifugal acceleration at $r = R$ gives
\begin{equation}
    \dfrac{\Delta p}{2\rho} = \upsilon^2.\label{eqn:cyclostrophic_balance}
\end{equation}

The dust devil's pressure gradient influences the ambient wind field and draws in air out to a distance $r = r_{\rm inf} = n R$, i.e. some number of radii out. If the ambient wind field has a lateral wind shear $\alpha \equiv \partial U/\partial x$, there will be a difference in velocity from one side of the devil to the other for the incoming air, $\Delta U \approx \alpha r_{\rm inf}$, which neglects factors of order unity. The attendant specific angular momentum $l$ can be estimated by multiplying this velocity difference by the lever arm $r_{\rm inf}$, i.e. $l \approx \alpha r_{\rm inf}^2$. Assuming this angular momentum is roughly conserved as the fluid travels from $r_{\rm inf}$ to $R$ implies $\alpha r_{\rm inf}^2 = \alpha n^2 R^2 \approx \upsilon R$ or 
\begin{equation}
    \upsilon \approx \alpha n^2 R.
\end{equation}
The appropriate value for $r_{\rm inf}$ (and therefore $n$) likely depends on the dust devil's properties and ambient conditions (e.g., wind shear, turbulent drag, etc.), but the exact dependence is unclear. Aside from assuming $r_{\rm inf} \gg R$ (previous studies have suggested $n = 4-10$ -- \citealp{2001JAtS...58..927R}), I leave it unspecified.

Using Equation \ref{eqn:cyclostrophic_balance}, we find
\begin{equation}
    R \approx \alpha^{-1} n^{-2} \left( \dfrac{\Delta p}{\rho} \right)^{1/2},\label{eqn:R_vs_Delta-p}
\end{equation}
again neglecting factors of order unity. 

Next, we can express the radius in terms of the dust devil height $h$. \citet{1998JAtS...55.3244R} suggested
\begin{equation}
    \Delta p = p_{\infty} \left\{ 1 - \exp \left[ \left( \dfrac{\gamma \eta}{\gamma \eta - 1}\right) \left(\dfrac{1}{\chi}\right) \left( \dfrac{\Delta T}{T_{\infty}}\right) \right] \right\}\label{eqn:Renno_Delta-p},
\end{equation}
where $\chi$ is the ratio of the gas constant $R_\star$ to the specific heat capacity at constant pressure $c_{\rm p}$ and is equal to 0.22 \citep{2000JGR...105.1859R}; $\gamma$ is the fraction of mechanical energy dissipated by friction near the surface; and $\Delta T$ the difference in temperature between the positive perturbation at the devil's center and the ambient temperature $T_\infty$. $\eta$ is the thermodynamic efficiency, given by 
\begin{equation}
    \eta = \dfrac{T_{\rm h} - T_{\rm c}}{T_{\rm h}}\label{eqn:Renno_eta},
\end{equation}{}
where $T_{\rm h}$ is the entropy-weighted mean temperature near the surface where heat is absorbed, and $T_{\rm c}$ is the same for the cold sink at the top of the dust devil. Estimates of $\eta$ based on field observations suggest $\eta \lesssim 0.1$ \citep[e.g.][]{2000JGR...105.1859R}. A useful approximation gives $T_{\rm h} \approx T_\infty$, while
\begin{equation}
    T_{\rm c} = \left[ \dfrac{ p_\infty^{\chi + 1} - p_{c}^{\chi + 1} }{\left( p_\infty - p_{c} \right) \left( \chi + 1\right) p_\infty^{\chi}} \right] T_{\rm h},\label{eqn:Tc}
\end{equation}
where $p_{\rm c}$ is the pressure near the top of the dust devil  \citep{2000JGR...105.1859R} and is related to the surface pressure as $p_{\rm c} \approx p_\infty \exp\left(-h/H\right)$ with $H$ the atmospheric scale height. For Mars, $H \ge 10\,{\rm km}$, and, although dust devils are sometimes observed that tall, usually they are a few km or less in height \citep{2008Icar..197...39S}. 

\begin{figure}
    \centering
    \includegraphics[width=\textwidth]{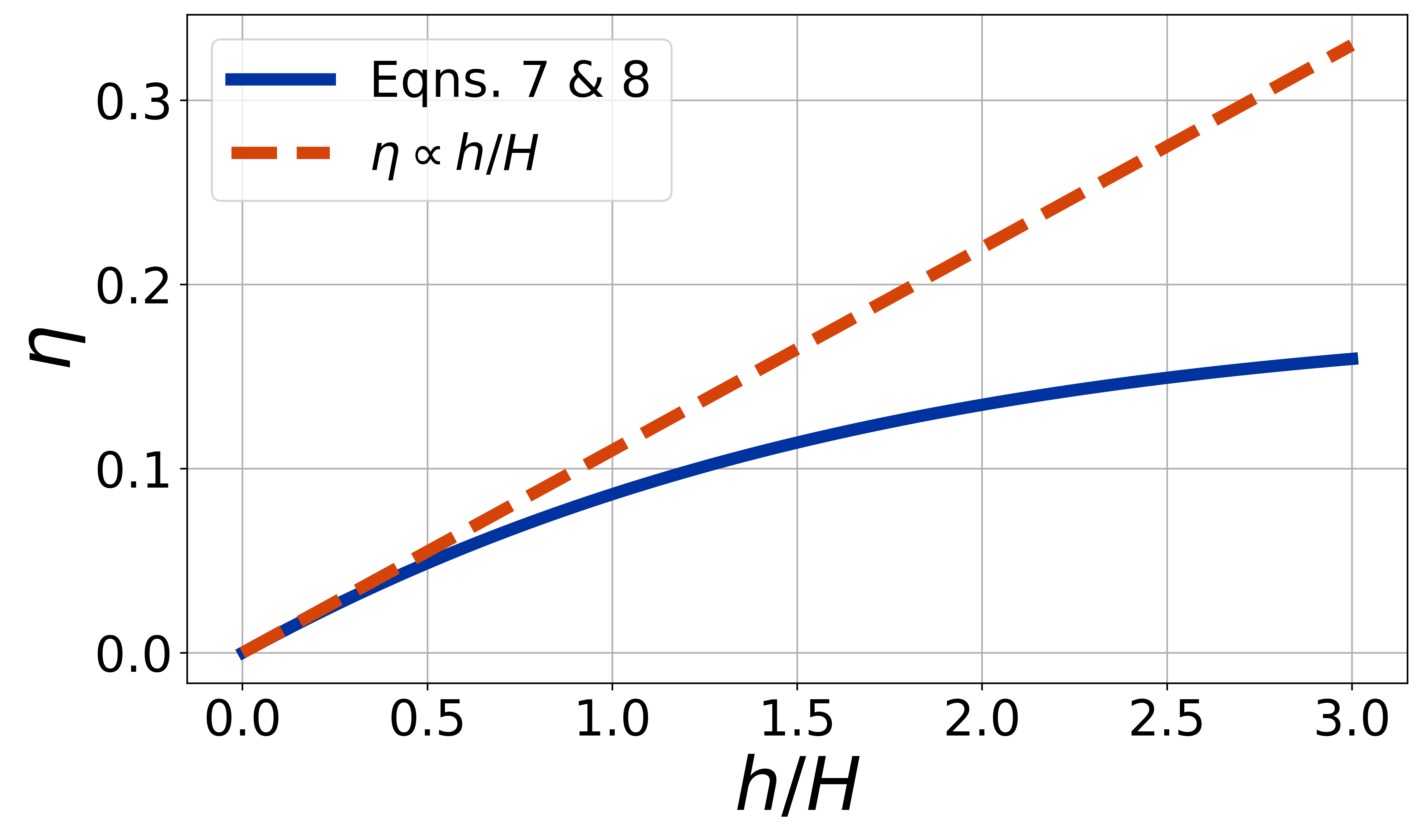}
    \caption{Dust devil thermodynamic efficiency $\eta$ as a function of dust devil height $h$ normalized to the atmospheric scale height $H$. The solid, blue line shows the full behavior given by Equations \ref{eqn:Renno_eta} and \ref{eqn:Tc}, while the dashed, orange line shows a linear approximation.}
    \label{fig:eta_vs_h-over-H}
\end{figure}

We can plug these expressions into Equation \ref{eqn:Renno_eta} and expand about small $h/H$:
\begin{equation}
    \eta \approx \frac{1}{2} \chi \left( \dfrac{h}{H} \right).\label{eqn:approx_eta}
\end{equation}
In other words, for most dust devils, the thermodynamic efficiency increases linearly with their heights. Figure \ref{fig:eta_vs_h-over-H} shows how $\eta$ depends on $h/H$ for a wide range of values and confirms the linear behavior for small $h/H$. We can plug Equation \ref{eqn:approx_eta} into Equation \ref{eqn:Renno_Delta-p} and again expand about small $h/H$:
\begin{equation}
    \Delta p \approx \left( \dfrac{\gamma R_\star \rho \Delta T}{2} \right) \left( \dfrac{h}{H} \right) \label{eqn:approx_Delta-p},
\end{equation}
with $p_\infty/T_\infty = R_\star \rho$.

Since $R$ depends on the scale of the pressure perturbation, which itself depends on $\eta$, we can write a relationship between $R$ and $h$ using Equation \ref{eqn:R_vs_Delta-p}:
\begin{equation}
    R \approx \alpha^{-1} n^{-2} \left( \dfrac{\gamma R_\star \Delta T}{H} \right)^{1/2}\ h^{1/2},\label{eqn:R_vs_h}
\end{equation}
with factors of order unity neglected.


Although Equation \ref{eqn:R_vs_h} provides a relationship between $R$ and $h$, it involves several parameters that are difficult to measure in practice. For instance, surveys of martian dust devils using space-based imagery (see Section \ref{sec:fitting}) can provide heights and radii, given sufficient resolution, but not $\alpha$ or $\Delta T$. However, we may expect that the unmeasured variables exhibit a range of values for any given $h$. With a sufficiently large population of dust devils, a model fit to the distribution of measured $R$- vs.~$h$-values (along with accurate uncertainties) should recover the underlying relationship. Indeed, as I show below, a fit to results from a dust devil survey closely resembles Equation \ref{eqn:R_vs_h}.

\section{Fitting the Model to Observational Data}
\label{sec:fitting}
\begin{figure}
    \centering
    \includegraphics[width=\textwidth]{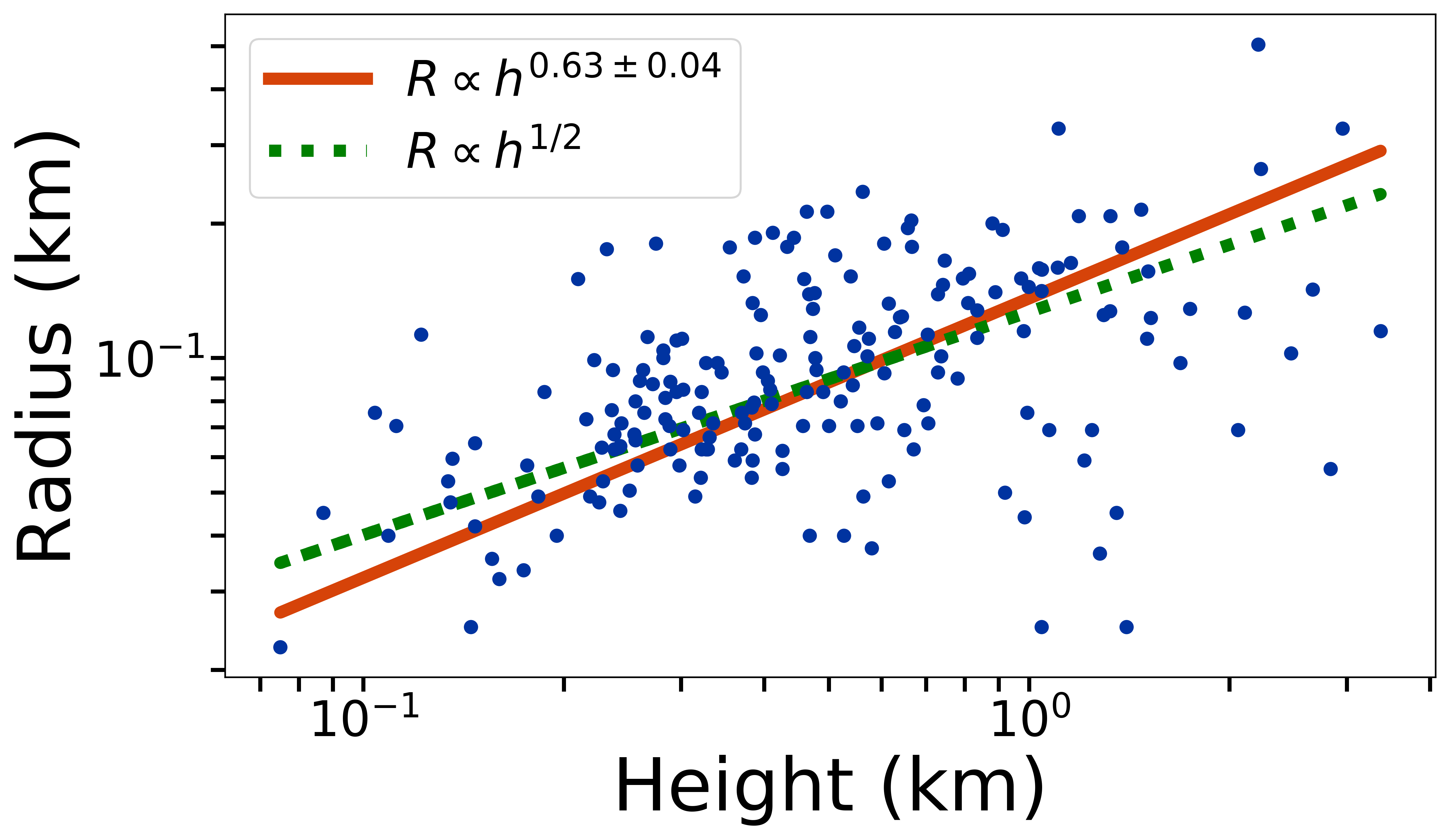}
    \caption{The blue dots are dust devil heights $h$ and radii $R$ in kilometers reported in \citet{2008Icar..197...39S}. The solid, orange line shows the line for which the best-fit exponent (0.63) is allowed to float, while the dashed, green line fixes the exponent at $1/2$ as in Equation \ref{eqn:R_vs_h}.}
    \label{fig:Fit_to_Stanzel_data}
\end{figure}

Numerous surveys involving space-based imagery have provided measurements of dust devil properties. The most voluminous survey, \citet{2006JGRE..11112002C}, reports more than 11k active devils imaged by the narrow- and wide-angle instruments of the Mars Global Surveyor's Mars Orbital Camera but only reports devil occurrence, not their radii and heights. Another comprehensive survey described in \citet{2008Icar..197...39S} provides estimates of diameters and heights for nearly 200 active devils using the Mars Express High Resolution Stereo Camera, with image resolutions between $12.5$ and $25\,{\rm m\ pixel^{-1}}$. The reported uncertainties on the diameters were typically $63\,{\rm m}$ and on the heights were typically $\ge 100\,{\rm m}$. I use these data, shown in Figure \ref{fig:Fit_to_Stanzel_data}, to test Equation \ref{eqn:R_vs_h}.

To fit these data, I applied two different models. For the first (shown as the solid, orange line in Figure \ref{fig:Fit_to_Stanzel_data}), I assumed $R \propto h^\Gamma$, with $\Gamma$ allowed to float. For the second fit, I fixed $\Gamma = 1/2$, as in Equation \ref{eqn:R_vs_h}. For both fits, I allowed the proportionality constant to float. Since both the ordinate and abscissa (radius and height, respectively) involve significant measurement uncertainties, I use the orthogonal distance regression algorithm, which can accommodate uncertainties along both dimensions \citep{odrref, scipy}, to fit the model parameters.

The best-fit $\Gamma = 0.63 \pm 0.04$ is $3.5\sigma$ discrepant from the value predicted by Equation \ref{eqn:R_vs_h}. This disagreement may arise from several factors. Most importantly, Equation \ref{eqn:R_vs_h} involves several important simplifying assumptions, including that $n$ is independent of ambient conditions and a dust devil's properties and that $h$ is independent of $\Delta T$. In reality, a larger ambient wind shear can drive enhanced turbulent dissipation \citep{arya1988}, potentially giving rise to an inverse relationship between $n$ and $\alpha$. We also expect a positive correlation between $\Delta T$ and $h$, although the level to which a convective plume rises also depends on the ambient lapse rate. In any case, the fact that the best-fit $\Gamma$-value closely resembles the predicted value suggests these effects are not significant.

The discrepancy may also arise from features of the survey itself. Although \citet{2008Icar..197...39S} give uncertainties for the diameters and heights, no details are provided regarding how they are determined, and so it is difficult to judge their accuracy. The exact value and uncertainty for $\Gamma$ depend sensitively on the measurement uncertainties. To demonstrate this dependence, I artificially doubled the uncertainties on the diameters (but not on the heights) and found that the best-fit $\Gamma$-value can be made to agree with $1/2$, meaning even a modest underestimate for the uncertainties can give discrepant results. Likewise the size of the surveyed population contributes to uncertainties on the model fit \citep{2015JGRE..120..401J}. By randomly selecting many different sub-sets of the reported diameter-height pairs half the size of the full survey, I find that I can often retrieve a best-fit $\Gamma$ consistent with $1/2$, meaning a larger survey might have given a different $\Gamma$-value. These analyses highlight the importance of a robust assessment of measurement uncertainties and of using the largest sample size possible when exploring dust devil population statistics.

\section{Conclusions}
\label{sec:conclusions}
Additional work can test the model presented here. Probes of active dust devils to explore internal structures and dust abundances, such as the work with instrumented drones described in \citet{2018RemS...10...65J}, would provide the most direct test of the scaling relationships described here. The arguments above also suggest a image survey to recover a larger population of dust devils with a detailed assessment of uncertainties could clarify the radius-height relationship. In fact, the population of dust devils identified but not measured in the survey reported in \citet{2006JGRE..11112002C} might be ideally suited.

The scalings here suggest other relationships that can be tested. For instances, combining Equations \ref{eqn:cyclostrophic_balance} and \ref{eqn:approx_Delta-p} allows us to estimate the eyewall velocity from a dust devil's height:
\begin{equation}
    \upsilon \approx \frac{1}{2} \left( \dfrac{\gamma R_\star \Delta T}{H} \right)^{1/2} h^{1/2},\label{eqn:velocity_vs_h}
\end{equation}
which, except for the scale height, is insensitive to ambient conditions (assuming they are suitable for dust devil formation). The momentum flux carried by a wind of speed $\upsilon$ scales as $\rho \upsilon^2$. Although the details of dust lifting can be complicated \citep[e.g.][]{1985wagp.book.....G}, once the grains are lifted, momentum conservation requires that their mass flux is proportional to the wind's momentum flux. The dust mass crossing an area oriented perpendicular to the flow in unit time is therefore proportional to $\upsilon^2$. This dust flux is transported around the circumference of the dust devil in an amount of time $\tau = 2\pi R/\upsilon$. Thus, at steady-state, the total dust mass transported around the eyewall is proportional to $\upsilon^2\tau = \upsilon R \propto h$. Of course, the actual dust content of a devil will also depend on the availability of dust in the region it forms, but with a large enough population, the underlying dependence on $h$ may be apparent. 

A more indirect test would be to compare the distribution of diameters measured by imagery surveys to the pressure profiles observed by martian landers. However, such an analysis may require a scheme to account for the biases of these lander surveys \citep{2018Icar..299..166J, 2019Icar..317..209K}. More challenging but perhaps enlightening might be measurements of ambient wind shear and its influence on dust devils \citep{arya1988}.

If future work can refine or improve the relationships presented here, dust devils may serve as probes of martian meteorology and dust cycle. For instance, Equation \ref{eqn:R_vs_h} shows that, given a measured height, a devil's radius depends on the atmospheric scale height. As the martian atmosphere heats and cools during the day, the scale height waxes and wanes, and so the radius-height relationship, as probed at different times of day, should measurably shift and constrain the near-surface heat budget \citep{Martinez2017}. Optical depth for devils with a given (or a narrow range of) properties may vary from region to region, depending on the availability of dust \citep{2002JGRE..107.5042B}, and provide input to models of the dust cycle.

\acknowledgments
This work was supported by grant number 80NSSC19K0542 from NASA's Solar System Workings program. The author acknowledges useful input from an anonymous referee.

\bibliography{main}
\bibliographystyle{aasjournal}



\end{document}